ERZION INTERPRETATION OF NEGATIVE PENETRATING COSMIC RAY PARTICLES EXCESS FLUX OBSERVED IN BUBBLE CHAMBER "SKAT"

Yu.N.Bazhutov
State Technical University (MADI), P.O.Box 169, 105077 Moscow, Russia
Bazhutov@hotmail.com / Fax: 7-095-151-0331

ABSTRACT. It is discussed the interpretation of negative penetrating cosmic ray particles excess flux observed in bubble chamber "SKAT" for the momentum range - $P > P_0 = 30$ GeV/c by Erzions, hypothetical heavy stable penetrating hadrons, proposed to explain the anomalous vertical muons energy spectrum at small depth underground. Here it is shown that negative charge of particles observed in "SKAT" is the same as predicted by theoretical Erzion model. The excess particles flux ($J \sim 10^{-5}$ cm$^{-2}$s$^{-1}$sr$^{-1}$) corresponds to the Erzion intensity observed by scintillation telescope in our previous experiment. The threshold momentum ($P_0$) and the track length threshold ($L_0 = 50$ cm of liquid BrF$_3$C) are in good accordance with Erzion stop path as for the single charged particle with mass - $M \cong 200$ GeV/c$^2$. But to don't contradict with all previous charge ratio results for cosmic ray muons in 30 – 100 GeV/c momentum range it is necessary to propose for such particles the Solar sporadic origin taking to account that both Erzion observations were in the active Sun years (April 23,1979 & July, 1999).

INTRODUCTION. 20 years ago to explain anomalous energy spectrum of vertical cosmic ray muons, observed at sea level and small depth underground (<100 m.w.e.) [1,2], it was proposed hypothesis of existing in cosmic rays new heavy stable penetrating hadrons [3]. From that time our experiments to search such particles were started [4,5,6]. Later the theoretical model $U(1) \times SU_1(2) \times SU_r(2) \times SU(3)$ of such particles (Erzions) has been created in framework of "mirror" models [7,8], which without contradictions to elementary particles Standard Model has explained large kind of another anomalous results in cosmic rays and nuclear physics [9-19]. At last after almost 20 years Erzions search they have been observed due to small vertical original scintillation telescope "Doch-4" [20,21,22]. The observed Erzions mass was $M_E = (175 +/- 25)$ GeV/c$^2$ and intensity at sea level – $J_E = (1.8 +/- 0.4) \cdot 10^{-6}$ cm$^{-2}$sr$^{-1}$s$^{-1}$ (at $E_E \leq 6$ GeV, $P_E \leq 50$ GeV/c$^2$). To confirm such Erzion discovery it was undertook the attempt of Erzions search on one of the largest bubble chamber (BC) "SKAT", exposed 16 years (1976-1992) on the neutrino beam of Serpukhov Proton Accelerator.

RESULTS INTERPRETATION. At the first step of such investigation it was found the excess flux ($J \cong 10^{-5}$ cm$^{-2}$sr$^{-1}$s$^{-1}$) for vertical ($\theta < 45°$) negative charged cosmic ray particles [23] in momentum range (30 GeV/c $\leq P \leq$ 126 GeV/c) in "SKAT" films from April 23, 1979 run (by liquid CF$_3$Br filled) and in December 00, 1991 run (by liquid C$_3$H$_8$ filled). At first let us analyze 1979 run. It was chosen only such events which were stared in BC ceiling and had track length more then $L_0 = 50$cm$= 76$g/cm$^2$. It is natural to suppose that $L_0$ is the stopped path for excess flux particles with minimum value of their momentum – $P_0 = 30$ GeV/c and all anomalous particles with $P < P_0$ will have their track length – $L < L_0$ and couldn't be among chosen events. If we suppose that anomalous particles are single charged particles and that they have only ionization energy loss we may use stopped path dependence from initial momentum for different mass of particles to compare with our results.
On the fig.1 you can see such dependence for different particles masses (100, 200, 1000 GeV/c$^2$) in liquid CF$_3$Br and the chosen events range, which corresponds to particles mass – $M \cong 200$ GeV/c$^2$ and correlates with our previous Erzion observation results [20-22]. Besides the "SKAT" excess flux ($J \cong 10^{-5}$ cm$^{-2}$sr$^{-1}$s$^{-1}$) for anomalous particles with momentum – $P > 30$ GeV/c roughly correlate with Erzion component intensity - $J_E = (1.8 +/- 0.4) \cdot 10^{-6}$ cm$^{-2}$sr$^{-1}$s$^{-1}$ (at $E_E \leq 6$ GeV, $P_E \leq 50$ GeV/c$^2$) for near momentum range [20-22]. And, at last, exactly negative but not positive charge was predicted by Erzion model [7,8].

DISCUSSION. The first that may cause real doubts here is large anomalous component intensity or the same it's large ratio (~30%) to muon's component in the same momentum range. But in this momentum range cosmic muon spectrum is studied more then 40 years very carefully the same as muon charged ratio [24-27]. To correlate these new results with all previous one can only if to propose their sporadic Solar origin with 11 years Solar activity variation. By the way 2 discussed here "SKAT" runs with the same excess flux for negative particles [23] were in active Solar years (1979 and 1991) so as Erzion observation by vertical original scintillation telescope "Doch-4" in 1999 [20-22]. Solar Erzion bursts can correlate with Solar proton bursts not every time, because nuclear absorption length for fast Erzions (E>20GeV) is about 100 time more than for fast protons (E>100MeV). Such Solar burst Erzions can be observed as excess flux ($J \sim 10^5$ $cm^{-2}sr^{-1}s^{-1}$) possibly only during active Sun years with real daily variation and with better registration them at small latitude because Earth magnetic field disperse them for such momentum range (50-100 GeV/c). Then only those Erzions, which entered in condensed atmosphere near vertical direction, have opportunity to pass it without absorption, because they are not relativistic for such energy and have large ionization losses. If they have energy less then critical value $E_0 \cong 12$ GeV, they are absorbed by atmosphere. Also Erzion is hadron (meson) [7,8], its nuclear interaction dose not play any role, because, at first, it is penetrating hadron [3] according to Additive Quark Model and, at second, it is not relativistic particle with large ionization loss. Negative charged Erzion – $E=\{U^{\wedge},d\}$ life time may be the same as for neutron ($\tau \sim 1000s$) that it can reach Earth from Sun without decay. Such way the problem of Erzion transportation may be decided. Next important problems are Erzion existence in Solar matter and its acceleration up to energy $E \sim 20$ GeV. The Erzions existence inside Solar matter has trivial decision in framework of Erzion Model [7,8] due to existence of absolutely stable neutral Erzion - $E^0=\{U^{\wedge},u\}$, as meson partner, from relict component or initial cosmic ray component. Both Erzion partners can transform each to another due to exchange nuclear reactions inside Solar matter. The last problem of Erzion acceleration is solved due to well-known Collective Accelerator Mechanism, in which massive particles of small concentration inside light particles collective are involved in collective moving of light particles with the same drift velocity. Such way light particles collective accelerate massive particles up to energy more then light particles energy by a factor of their mass ratio. So in the proton flux with energy - $E_P = 100$ MeV must be Erzions with energy of $E_E = 20$ GeV, because the Erzions concentration in Solar matter may be very small ($C_E \sim 10^{-15}$). Due to their nuclear penetrating peculiarity their flux may be reached at the exit from condensed Solar matter by some orders. Such way due to this Collective Accelerator Mechanism our rough estimation at first step have showed that observed in our experiments [20-23] Erzions flux ($J \sim 10^{-5}$ $cm^{-2}sr^{-1}s^{-1}$) can be explained.

CONCLUSION. Of course, the author understands that all explained here is only first rough estimations. It is needed for more correct and care calculations in future. But I wanted to show in this article how may be rich and prosperous Erzion model for interpretation inconceivable at first view experimental results and phenomenon and I hope that it is real idea. Now experiments for Erzions search are continued on 3 Russian installations: on the automatic modified scintillation telescope "Doch-4a" together with colleagues from RNC "Kurchatovskiy institute", it is continued review of films from the bubble chamber "SKAT" together with SSC Institute of High Energy Physics colleagues and it is starting on the large Muon Telescope of Moscow Physical Engineering Institute.

ACKNOLEDGMENTS. I want gratefully thank for continuous help and support in Erzion researches my scientific teachers and colleagues: Drs. G.B.Khristiansen, B.A.Khrenov, L.A.Mikaelyan, G.T.Zatsepin, V.S.Murzin, G.M.Vereshkov, V.P.Koretsky and all other connected with me by this interest problem.

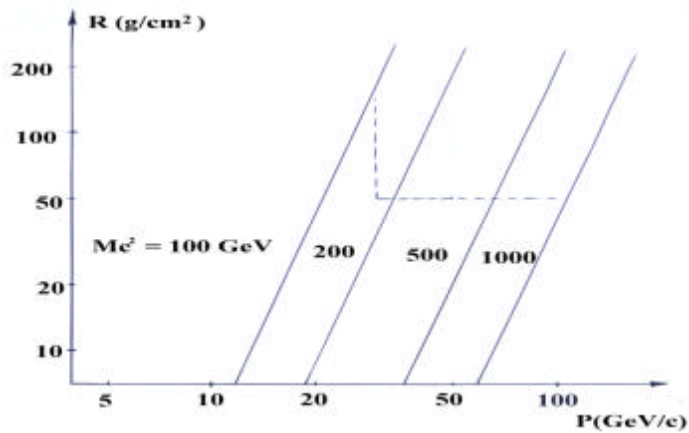

Fig.1. Dependence of stop path from start momentum for different particle masses in liquid $CF_3Br$.